\documentclass[preprint,showpacs,amsmath,amssymb]{iopart}

\pdfoutput=1
\usepackage{graphicx}
\usepackage[pdftex,dvips]{epsfig}

\begin{document}

\title[Electronic structure of graphene flakes near the Fermi level]
{Electronic structure of triangular, hexagonal and round graphene 
flakes near the Fermi level}

\author{H P Heiskanen, M Manninen, and J Akola}

\address{Nanoscience Center, Department of Physics, P.O.Box 35, 
FI-40014 University of Jyv\"askyl\"a, Finland}

\ead{matti.manninen@jyu.fi}

\date{\today}

\begin{abstract} 

The electronic shell structure of triangular, hexagonal and round 
graphene quantum dots (flakes) near the Fermi level has been studied 
using a tight-binding method. The results show that close to the 
Fermi level the shell structure of a triangular flake is that of free 
{\it massless} particles, and that triangles with an armchair edge 
show an additional sequence of levels (``ghost states"). These levels
result from the graphene band structure and the plane wave solution 
of the wave equation, and they are absent for triangles with an 
zigzag edge. All zigzag triangles exhibit a prominent edge state at 
$\epsilon_F$, and few low-energy conduction electron states occur both 
in triangular and hexagonal flakes due to symmetry reasons. Armchair
triangles can be used as building blocks for other types of flakes 
that support the ghost states. Edge roughness has only a small effect 
on the level structure of the triangular flakes, but the effect is 
considerably enhanced in the other types of flakes. In round flakes, 
the states near the Fermi level depend strongly on the flake radius, 
and they are always localized on the zigzag parts of the edge.

\end{abstract}

\maketitle

\section{Introduction}

Nearly free electrons trapped by a high-symmetry potential exhibit a 
shell structure that arises from the symmetry-induced degeneracy and 
bunching of energy levels of different radial modes. Such level 
structure has been observed in metallic clusters and semiconductor 
quantum dots (for reviews see \cite{deheer1993,reimann2002}). Usually, 
the shell structure is associated with a spherical or circular symmetry, 
but it exists also, for example, in three-dimensional icosahedral
\cite{mansikkaaho1992} and two-dimensional triangular clusters
\cite{reimann1997}. The shell structure is a single-particle property 
and can be understood on the basis of the jellium model of delocalized 
electrons \cite{ekardt1984} or the tight-binding approach \cite{manninen1991}.

In two-dimensional systems, the most interesting confinement 
geometries for electrons are a circle, hexagon, and triangle. Obviously, 
the circle has the highest symmetry of these and the triangle the lowest.
Surprisingly, however, it is the triangle that has the most persistent 
shell structure and also a regular supershell structure \cite{brack1997}. 
The triangular shape is preferred in two-dimensional metallic systems
\cite{reimann1997,kolehmainen1997,janssens2003}, in plasma clusters
\cite{reimann1998}, and it is observed also in semiconducting silicon
clusters \cite{lai1998}. 

The shell structure of quantum dots and metal clusters is caused by nearly 
free conduction electrons. In the case of graphene, the situation is 
different due to the peculiar band structure. The Fermi surface consists 
of a set of discrete points, and the electron (hole) dispersion relation 
of the conduction (valence) band is linear. Recent experiments have shown 
that nanometer-sized graphene flakes can be produced on various surfaces 
\cite{berger2004,novoselov2004,novoselov2005,berger2006,novoselov2007,geim2007,li2007},
which has induced a significant amount of theoretical interest
\cite{alicea2005,gusynin2005,tworzydlo2006,gusynin2006,zhou2006,yamamoto2006,nomura2007,son%%@
%%@
2007,areshkin2007,chen2008,castro2008,akhmerov2008}.

In this article, we show that finite graphene flakes (or quantum dots) 
have an interesting energy spectrum close to the Fermi level. The most 
common edges of graphene are the so-called armchair and zigzag edges. It 
turns out that the energy spectrum of graphene flakes depends strongly on 
the type of the edge, and that flakes of similar size and shape can exhibit
distinctly different electronic structure (selection rules). In an earlier 
report \cite{akola2008}, we reported results for triangular graphene flakes 
and showed that a simple tight-binding (TB) model that considers only the 
carbon $p_z$ electrons produces a similar shell structure than a full 
electronic structure calculation with all the valence electrons (based on 
the density functional theory, DFT). Moreover, the results showed that the 
electronic levels close to the Fermi energy can be understood as those of 
free massless electrons confined in a triangular cavity. Herein, we shall 
further investigate the peculiarities of the graphene electronic structure 
that are caused by the geometry and edge structure of the flake. 

\begin{figure}[h]
\hfill\includegraphics[width=0.7\textwidth]{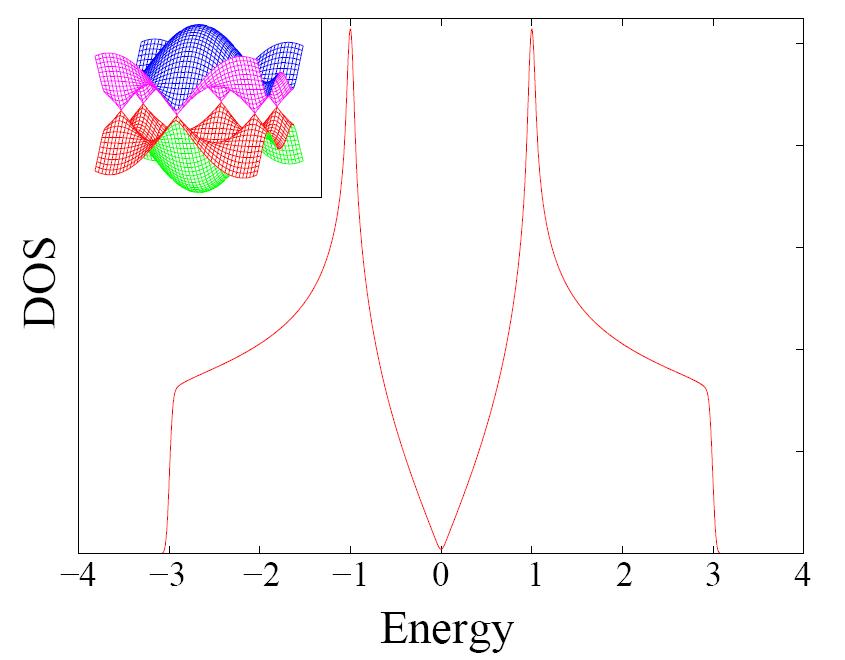}
\caption{The density of states ($p_z$ electrons) of an infinite graphene 
sheet for the TB method used. Inset: cross-over of the valence and 
conduction bands at the Fermi energy.}
\label{bands}
\end{figure}

The atomic $p_z$ electrons perpendicular to the graphene plane are 
responsible for the band structure shown in figure \ref{bands}, where the 
valence and the conduction bands meet at the corners of the hexagonal 
Brillouin zone \cite{wallace1947,elliot1998}. The Fermi surface consists 
of a discrete set of these points of high-$k$ value, and the resulting 
density of states (DOS) has a zero weight (as well as zero band gap) at 
the Fermi energy $\epsilon_{\rm F}$. The dispersion relation is linear in 
the near vicinity of the Fermi level. Since the atomic $p_z$ electrons are 
perpendicular to the graphene plane their interaction with the neighboring 
atoms does not have any directional dependence and the TB model can be 
reduced to the traditional H\"uckel model
\begin{equation}
H_{ij}=\left\{\begin{array}{rl}
-t, & {\rm if} \quad i,j \quad{\rm nearest \quad neighbours}\\
0, & {\rm otherwise}\\
\end{array}\right. 
\end{equation}
where the hopping parameter $t$ (resonance integral) determines the width 
of the bands and the on-site energy is chosen to be $\epsilon_F=0$. 
We present our results in units $t=1$ (in real graphene our unit $t$ 
corresponds to about 2.6 eV). It is important to note that the simple 
TB model becomes equivalent to that of the free electron model when the 
electron wave length becomes much larger than the interatomic distance
\cite{manninen1991}. This is valid at the bottom of the valence band 
where the free electron model gives the correct shell and supershell 
structure \cite{akola2008}. The situation is more complicated near the 
Fermi level where the electron wave length ascribes to the interatomic 
distance. However, as we shall see, also there the level structure can 
be understood in terms of the free electron model, but now for 
{\it massless} electrons.

In the following, we consider graphene flakes that are cut out from a 
perfect infinite graphene sheet and neglect the effects of the substrate 
as well as the passivation of dangling bonds. The passivation, say with 
hydrogen, involves $sp^2$ hybridized orbitals and is expected to have 
only a marginal effect on the perpendicular $p_z$ electron states
\cite{wallace1947,elliot1998}. This approximation was supported by our earlier 
work where we compared the full DFT calculations of hydrogen passivated 
graphene flakes with the results of the simple H\"uckel model without
passivation \cite{akola2008}. Note however, that our simple model can not 
account for possible spin-polarization of the edge states with large %%@
degeneracy\cite{son2006}.

\section{Triangular graphene flakes}

The Fermi level of graphene consists of two equivalent points at the 
border of the Brillouin zone (see figure \ref{bands}) where the 
conduction and valence bands open as circular cones resulting in a 
linear dispersion relation for electrons $\epsilon({\bf k})=C\hbar k$, 
where $C$ is the velocity. Thus, it is to be expected that the 
electron dynamics is not determined by the Schr\"odinger equation but 
by the equation of massless particles (Klein-Gordon or Dirac equation). 
The simple wave equation for a triangular cavity has an analytic solution
\cite{borghis1958} which gives the energy eigenvalues
\begin{equation}
\epsilon_{n,m}=\epsilon_1\sqrt{n^2+m^2-nm},
\label{levels2}
\end{equation}
where $m$ and $n$ are positive integers with $n\ge 2m$.
The state with $n=2m$ is nondegenerate while states with
$n>2m$ have a degeneracy 2. In our case
$\epsilon_1=2\pi t/\sqrt{3N}$, $N$ being the number of atoms.
In the case of Schr\"odinger equation (i.e. electrons with mass),
the exact solution gives $\epsilon \propto n^2+m^2-nm$, i.e. 
Eq. (\ref{levels2}) without the square root. It is interesting to 
note that the exact solution for the wave equation was presented by 
Lame already in 1852 \cite{lame1852}, as noted by Krishnamurthy
who studied the corresponding solution of the Schr\"odinger equation
\cite{krishnamurthy1982}. The corresponding wave functions can 
be found in \cite{doncheski2003}. 

The eigenvalues of Eq. (\ref{levels2}) are solutions of the wave equation
for massless particles, for example, for elastic waves, for electromagnetic waves
or for the positive energy solutions of the Klein-Gordon equation. 
We want to emphasize that we have not shown that they are solutions
of the Dirac equation where the boundary conditions are tricky for a 
cavity\cite{tworzydlo2006,akhmerov2008,castroneto2008}. However, our numerical
solutions of the TB problem for large triangular flakes are in excellent 
agreement with those of Eq. (\ref{levels2}).

The electronic density of states (DOS) of a finite system (flake) 
consists of a set of discrete energy levels. Instead of plotting the 
level structure it is more useful to study the density of levels since it points 
out more clearly the exact and nearly exact degeneracies of levels as well as 
the shell structure, which manifests itself as a regular variation of the level density.
It is thus useful to define a continuous DOS by using a Gaussian convolution 
of the discrete levels $\epsilon_i$:
\begin{equation}
g(\epsilon)=\frac{1}{\sigma\sqrt{2\pi}}\sum_i e^{-(\epsilon-\epsilon_i)^2/2\sigma^2},
\end{equation}
where $\sigma$ is the width of the Gaussian.

\begin{figure}[h]
\hfill\includegraphics[width=0.8\textwidth]{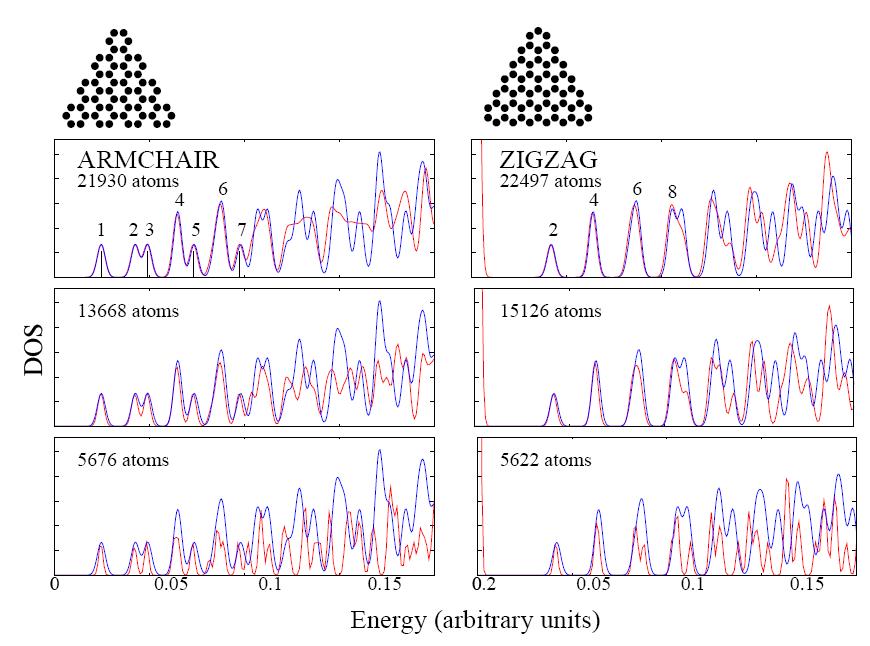}
\caption{TB-DOS above the Fermi energy 
for triangular flakes (red curves), compared to the DOS of equation 
(\ref{levels2}) (blue curves). 
The sizes of the triangles are given as numbers of atoms.
Note that the triangles showing the geometries are much smaller.
The energy is in units 
of $t$ for the largest armchair and zigzag triangles, respectively. 
For the smaller sizes the energy has been scaled by the square root 
of the number of atoms in order to get the peaks at the same 
positions.}
\label{dos2}
\end{figure}

Figure \ref{dos2} shows TB-DOS above the Fermi energy for three 
graphene triangles ($\sim$22000, 15000 and 5600 atoms) with armchair 
and  zigzag edges and compares them with the DOS of free massless 
electrons (see Eq. (\ref{levels2})). For armchair flakes, the 
comparison includes now additional (forbidden) index values $m=n$. 
The results are the following: 
(i) Each energy level has an additional degeneracy of two due 
to the two equivalent points at $\epsilon_F$. 
(ii) The zigzag triangle shows the levels of equation (\ref{levels2}) 
with index values $m\ge 1$ and $n\ge 2m$ while the armchair edge 
shows also those where $n=m$. We call these additional states (where 
$n=m$) as ``ghost states".
(iii) Equation (\ref{levels2}) describes only the lowest states 
accurately and is more successful for larger triangles. 
(iv) Due to the sparseness of the states, supershell oscillations of 
the massless particles become visible only in the large triangles 
(although they are visible at the bottom of the band already in small 
triangles \cite{akola2008}). 
(v) The zigzag edge supports particularly visible edge states 
\cite{kobayashi1993,nakada1996}
that appear at $\epsilon_F$  as a prominent peak (figure \ref{dos2}). 
The number of these states equals the number of the outermost edge 
atoms in zigzag triangles, which is $N_{\rm ss}=\sqrt{N}$.
We shall return to the edge states in section \ref{roundflakes}.

\begin{figure}[!ht]
\hfill\includegraphics[width=0.8\textwidth]{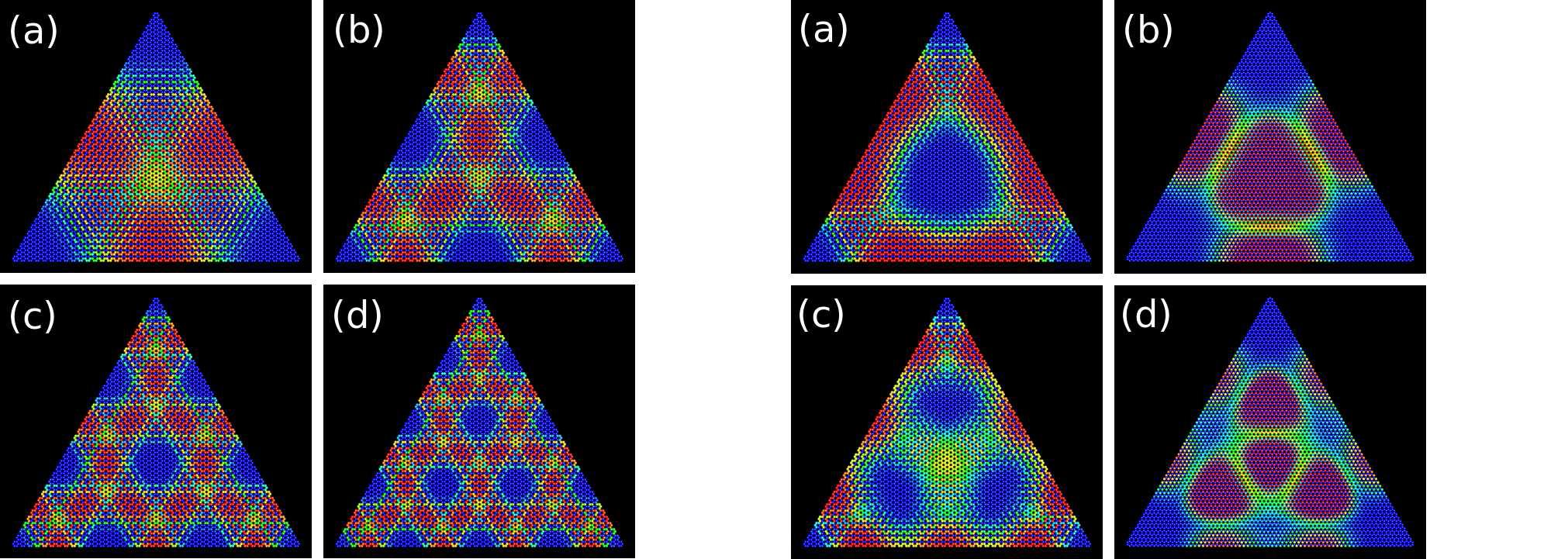}
\caption{LEFT FOUR: Electron density of (a) the 1st, (b) 3rd, (c) 5th, 
and (d) 7th energy levels above the Fermi energy in armchair triangles 
(``ghost states", labeled in Fig. \ref{dos2}). Each level has a 
degeneracy of two. RIGHT FOUR: Electron density of ((a) and (b)) the 2nd
and ((c) and (d)) the 4th energy levels above the Fermi energy (labeled 
in Fig. \ref{dos2}) for armchair and zigzag triangles of 4920 and 5181 C 
atoms, respectively. Color scale from blue to red, blue corresponds to 
vanishing density. Each figure shows the sum of the densities of the 
degenerate states.}
\label{ghost}
\end{figure}

Graphene ribbons with armchair edge show density of states with or without
a gap, depending of width of the 
ribbon\cite{nakada1996}. In the case of triangular flakes with armchair edge
no such effect was seen. The only size dependence observed was 
the scaling of the energy levels with the flake size.

The lowest conduction states that are numbered in figure \ref{dos2} show 
fascinating details, and the electron densities of such states are 
visualized in figure \ref{ghost}. The above-mentioned ghost states (left,
labeled by odd indices) show an interesting feature as they have a simple 
geometric pattern of triangular symmetry. The size (number) of the 
triangles decreases (increases) with increasing energy, i.e. the pattern 
repeats itself. These ghost states are completely absent for the zigzag 
triangles, and they correspond to quantum numbers of Eq. (\ref{levels2}) 
not allowed for free electrons in a triangular box (i.e. $n=m$ (with 
extra degeneracy) in Eq. (\ref{levels2})). Previously, we calculated the 
same states for a smaller armchair triangle with a DFT method (330 C atoms, 
60 passivating H atoms) \cite{akola2008}. The internal structure (symmetry) 
of the states was clearly similar, and therefore, the phenomenon is 
independent of the triangle size and the model used. We shall discuss the 
ghost states in detail in Section \ref{gstates}.

Figure \ref{ghost} (right) shows the electron densities corresponding to 
the ``normal" low energy states that obey the standard selection rules 
($m\ge 1$ and $n\ge 2m$). Again, the electron density does not necessarily 
vanish at the edges of the triangle. Interestingly, the corresponding states 
for the armchair and zigzag triangles (with the same energy and quantum 
numbers $n$ and $m$) display nearly an anticorrelation: The maxima in zigzag 
triangles are minima in armchair triangles and vise versa. Overall, the 
states close to the Fermi level appear very different from those at the 
bottom of the band. They are {\it not} simple densities of massless 
particles confined in a triangular cavity since the density profile does 
not decay to zero at the edges. The corresponding electron levels are close 
to the Brillouin zone boundary, having large $k$-values, and the wave 
functions have pronounced oscillations with wave lengths that are related to 
the hexagonal unit cell size. These oscillations enable that the wave 
function can be formally zero at the edges, but the corresponding pseudowave 
function of the massless particle does not necessarily vanish. 

\section{Hexagonal graphene flakes}
\label{hexaflakes}

Similar TB calculations were performed for hexagonal graphene flakes with
armchair and zigzag edges. The comparison between hexagonal and triangular 
flakes is based on hexagons that were cut from the corresponding triangles 
(taking the corners off). In general, the level structure is more complicated 
but some similarities with the triangular flakes can be found. We observe the 
following results:
(i) The zigzag edge supports edge states (see section \ref{roundflakes}) 
while the armchair edge results in a gap at $\epsilon_F$. 
(ii) The electron densities of the states {\it near} $\epsilon_F$ display 
the main amplitude at the edges/corners both for the zigzag and armchair 
edges. 
(iii) For the zigzag flake, the number of states near the Fermi level 
depends on the size of the flake.
(iv) Hexagonal flakes display few states that have exactly the same electron 
density as the original triangles (cutting off the corners). 
(v) In most cases, the electron densities are different than in the 
corresponding triangular flakes or hexagons with the other type of edge. 
It is also worth mentioning that at the bottom of the valence band the 
($p_z$) electrons act as free particles not seeing the atomic lattice,
which is a case similar to triangles. This makes the supershell structure 
visible at the bottom of the valence band, but it is not as clear as in the 
triangular graphene flakes. 

\begin{figure}[ht]
\hfill\includegraphics[width=0.8\textwidth]{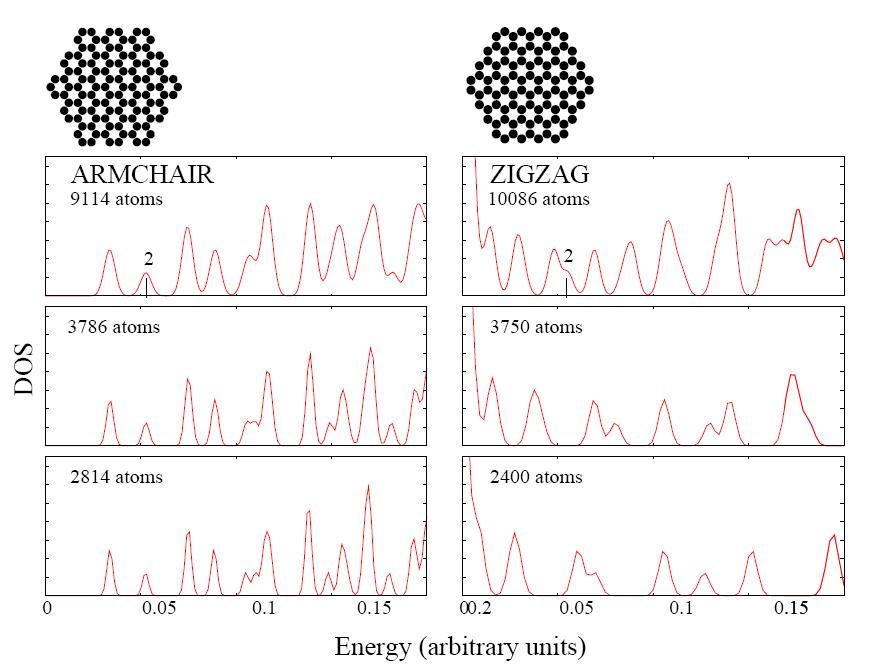}
\caption{TB-DOS of the hexagonal flakes with armchair (left panel) and 
zigzag (right panel) edge. The energy is in units of $t$ for the largest 
hexagons. For the smaller sizes the energy has been scaled by the 
square root of the number of atoms (scaling factor = $\sqrt{N_1/N_2}$, 
where $N_1$ is the number of atoms in the flake 1 and $N_2$ atoms in the 
flake 2). The geometries are shown as small hexagons.}
\label{hexdos}
\end{figure}

\begin{figure}[!ht]
\hfill\includegraphics[width=0.8\textwidth]{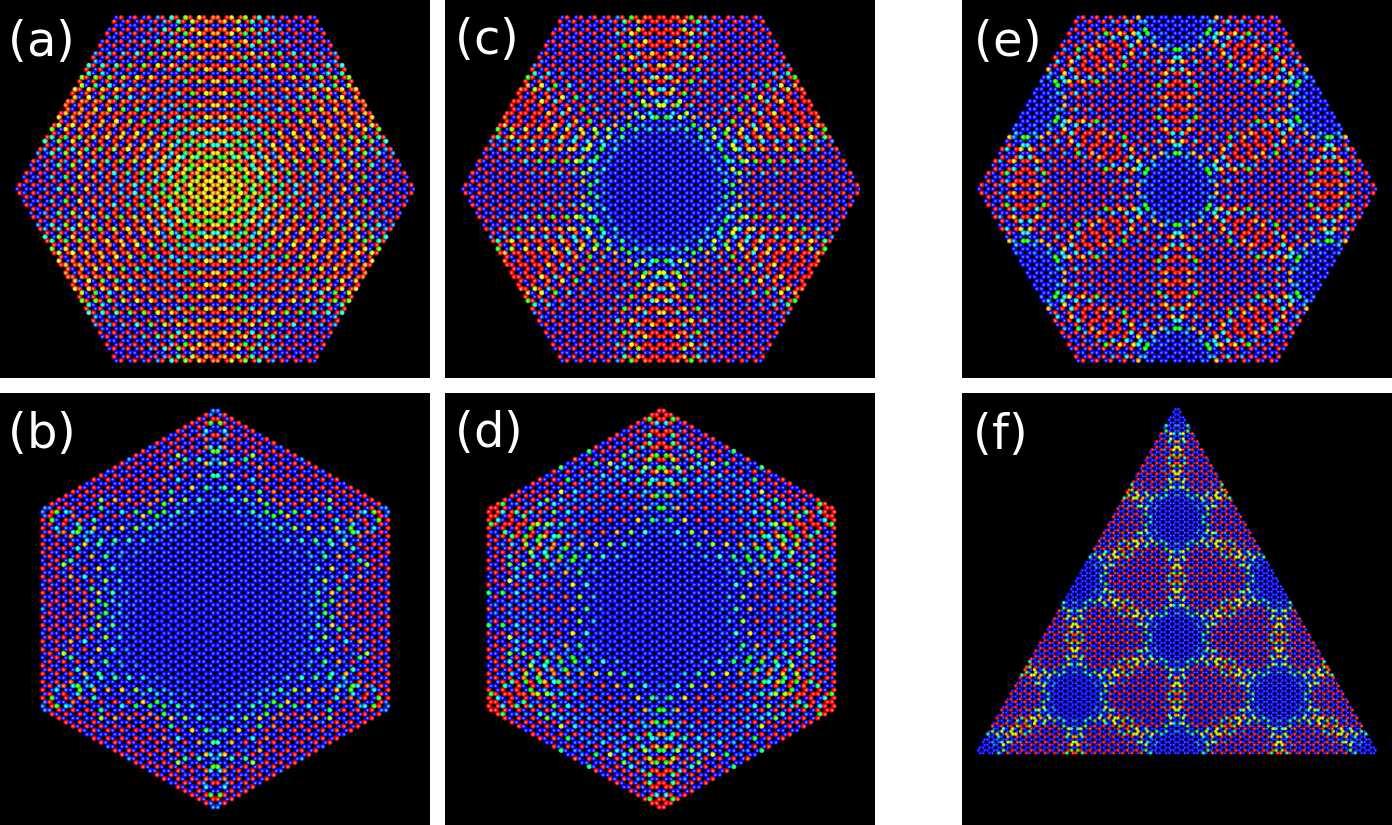}
\caption{Electron density corresponding ((a) and (b)) the first peak before 
the peak ``2" in Fig. \ref{hexdos}) and ((c) and (d)) the peak ``2" for the 
armchair (upper) and zigzag (lower) hexagons. (e) and (f) show a state 
that occurs both in the armchair-edged hexagon and triangle (peak ``6" 
in Fig. \ref{dos2}). Color scale from blue to red, blue corresponds to 
vanishing density.}
\label{hexdens}
\end{figure} 

%%% Number of atoms in the previous figure???

Figure \ref{hexdos} shows the DOS of hexagonal flakes. In the armchair 
panel (right), the DOS has been scaled by size in order to get the peaks 
to coincide near the Fermi level. The scaling factor $\sqrt{N_1/N_2}$ is 
the same as in the case of triangles. In the zigzag flakes, the scaling 
does {\it not} bring the peaks at the same positions. This is a special 
feature that does not exist in triangular flakes. In general, the zigzag
flakes have a peak and the armchair flakes display a gap at $\epsilon_F$, 
which is the case for triangles also. The hexagonal armchair flakes display 
states near the Fermi level that are in a sense universal: they do not 
depend on the size of the flake (cf. triangles, figure \ref{dos2}). The 
armchair flakes do not exhibit any states that could be regarded as ghost 
states suggesting that these are characteristic for the armchair triangles 
only. However, as will be shown in Section \ref{gstates}, a slight 
modification of the hexagonal flakes changes the situation. We also note 
that the DOS near the Fermi energy has some peaks that coincide with the 
ones of the triangles, and there is one state that is common in all the  
triangular and hexagonal flakes: the peak ``2" in figures \ref{dos2} and 
\ref{hexdos}. Furthermore, the states that have exactly the same energies 
in armchair triangles and hexagons display similar electron densities due 
to the common symmetry properties (figure \ref{hexdens}(e) and (f)). 

\section{Ghost states}
\label{gstates}

\begin{figure}[!ht]
\hfill\includegraphics[width=0.8\textwidth]{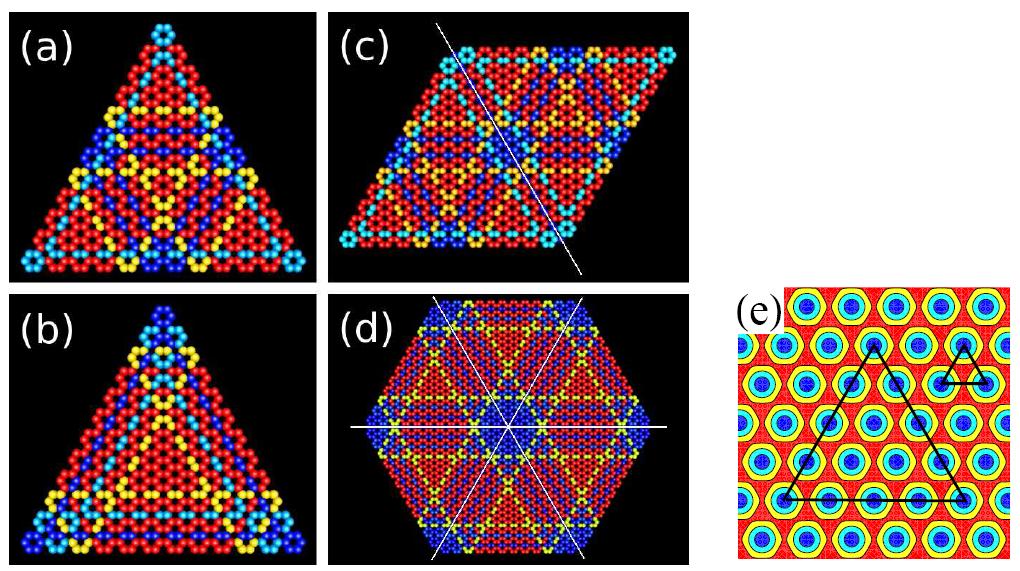}
\caption{Ghost states in ((a) and (b)) armchair triangles, and the 
corresponding ghost states (2nd and 1st) in (c) an imperfectly built 
rhombus and (d) hexagon. The extra rows of atoms are marked with the 
narrow white lines. (e) shows a standing wave solution of the wave 
equation in an infinite two-dimensional (hexagonal) system. The small 
and large triangles in (e) demonstrate how the ghost states (1st and 
4th) appear in triangular flakes.}
\label{ghostflakes}
\end{figure}

The hexagonal armchair flakes of section \ref{hexaflakes} do not exhibit 
the peculiar ghost states. The reason is that our flakes obey the armchair 
construction exactly: the corners are those of a perfect honeycomb pattern. 
However, the ghost states will re-appear if the flakes are built differently. 
This can be understood by studying a triangular armchair flake with ghost 
states (figures \ref{ghostflakes}(a)-(b)). The boundary conditions of the 
tight-binding problem require that the wave function is zero at the 
(imaginary) lattice sites just outside the triangle. Now, we can put two 
triangles together as in the rhombus-shaped flake shown in Fig. 
\ref{ghostflakes}(c)), and add an additional row of atoms between the 
triangles. This system has naturally the same ghost states as the 
original triangle. Similarly, we can construct hexagonal flakes with ghost 
states as shown in Fig. \ref{ghostflakes}(d), and it is clear that any shape 
consisting of equilateral triangles can exhibit ghost states. The only 
requirement is that an additional row of lattice sites (atoms) is added at 
the interface of the triangles. The ghost states in different triangles are 
then completely decoupled although they appear as continuous wave functions, 
and the wave function is exactly zero at the interface. This is also the 
reason why the ghost state pattern repeats itself: The high-index (large 
$n=m$) ghost states are the same in large triangles as the low-index ghost 
states in small triangles, and the energy is exactly the same if the side 
length $L$ of the large triangle is commensurate with that of the small 
triangle. 

At this point, it is important to note that the hexagonal flakes 
constructed according to the prescription above do not have perfect 
corners. Instead of the armchair edge just bending over, they have a small 
region of zigzag edge at the corners. Similarly, the extra row of atoms in 
the rhombus shown in figure \ref{ghostflakes} results in that the corners 
do not follow the armchair construction. The ghost states disappear, if 
the rhombus is made by merging two triangles together without an additional 
row of atoms.

The appearance of ghost states in the TB model reflects the balance between 
the graphene band structure and the free electron states in two-dimensional 
systems. Figure \ref{ghostflakes} (e) shows the combined density of two 
degenerate plane waves for free electrons, which can be expressed as
\begin{equation}
n(x,y)=\sin^2{qx}+\sin^2(\frac{1}{2}qx+\sqrt{3}qy)+\sin^2(\frac{1}{2}qx-\sqrt{3}qy).
\end{equation}
This density has a clear similarity of that of the ghost states, except that 
the rapid oscillations from atom to atom are absent. Again, we want to remind 
that the (pseudo) wave function of these ``Dirac fermions" above the Fermi 
energy does not need to be zero at the edge of the triangular cavity since 
the rapid oscillations take care of this boundary condition. Thus, also 
solutions where the derivative of the pseudowave function is zero are 
allowed. 

\section{Edge roughness}

The effect of edge roughness was studied for triangular and hexagonal flakes. 
We removed randomly 10, 38, or 50  percent of edge atoms and studied how it 
affected the DOS and electron densities of the states near $\epsilon_F$. The
atom removal process avoided situations where the possible remaining atom had 
only one nearest-neighbour, and such atoms were taken out. 

\begin{figure}[!h]
\hfill\includegraphics[width=0.7\textwidth]{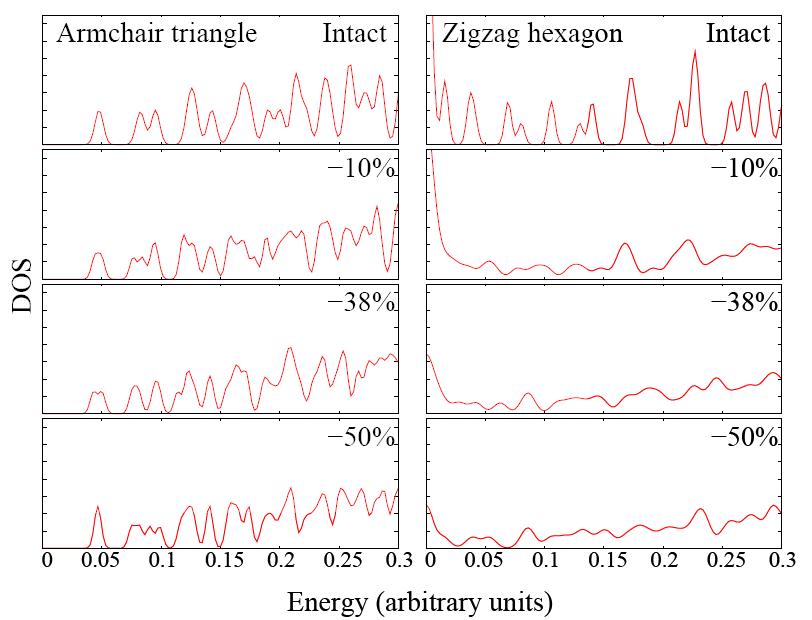}

\hfill\includegraphics[width=0.7\textwidth]{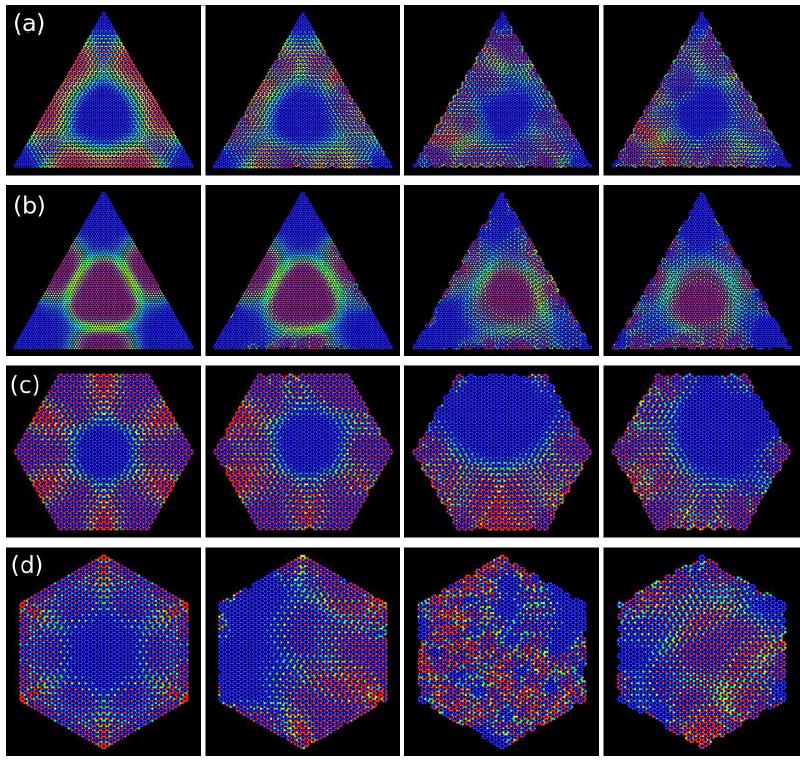}
\caption{UPPER PANELS: Effect of the edge roughness on TB-DOS for an 
armchair triangle (left) and zigzag hexagon (right).
LOWER PANELS: Effect of the edge roughness on the electron density
of the 2nd conduction electron state for armchair and zigzag flakes. 
Triangular flake with (a) armchair edge and (b) zigzag edge. Hexagonal 
flake with (c) armchair edge and (d) zigzag edge. From left to right: Intact 
edge, 10, 38, and 50 percent of the edge atoms removed. Color scale from blue 
to red, blue corresponds to vanishing density.}
\label{roughdos}
\end{figure}

The results are collected in figure \ref{roughdos} which shows the TB-DOS 
above the Fermi energy (upper panel) and the electron density of the 2nd 
conduction electron state (peak ``2", lower panel). Especially in the case of 
hexagonal flakes, the edge roughness has a noticeable effect on DOS and 
electron densities. Already a small edge roughness causes the degeneracy of 
the states to break up, and for example, removal of only 10\% of edge atoms in 
the zigzag-edged hexagonal flake results in a significant perturbation, and 
the electron density pattern of the intact flake cannot be identified anymore.
The changes are less dramatic for triangular flakes, and the pattern of the
2nd conduction state is always recognizable. The DOS curves indicate that
the states closest to $\epsilon_F$ are the most robust against edge 
roughness. Finally, the electron density of the corresponding states seems 
to avoid the rough parts of the edge in the case of armchair edge and favor 
them in the case of zigzag edge. 

\section{Round graphene flakes}
\label{roundflakes}

\begin{figure}[!h]
%\hfill\includegraphics[width=\textwidth]{figure9upper.eps}
\hfill\includegraphics[width=\textwidth]{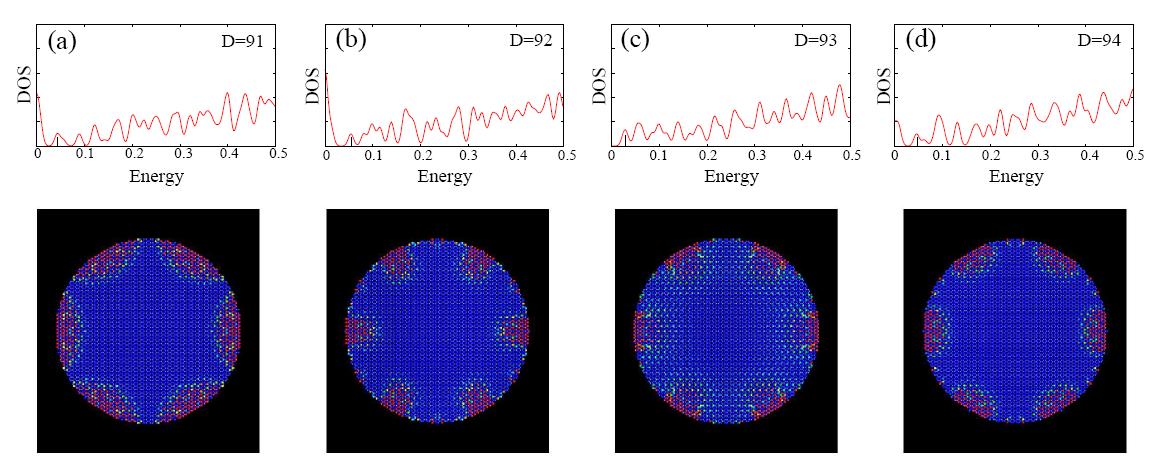}
\caption{TB-DOS of round graphene flakes (upper panel) and electron 
densities (lower panel) of the first conduction electron state above $\epsilon_F$ 
(marked with a tic in the DOS panels). The flake diameters (D) are (left 
to right): 91, 92, 93, and 94 times the nearest-neighbour distance
corresponding to 4980, 5118, 5238, and 5338 atoms, respectively. Color 
scale from blue to red, blue corresponds to vanishing density.}
\label{roundfig}
\end{figure}

Finally, we have studied round (circular) graphene flakes. Round flakes
were cut out of a graphene sheet as follows: The center was chosen to 
be a high-symmetry point (an atom or a center of a hexagon), and all the 
atoms inside a chosen radius were included. After this, all the edge 
atoms with only one nearest-neighbour were removed. The number of atoms 
is thus determined by the chosen radius and center. Figure \ref{roundfig} 
shows the TB-DOS above the Fermi level for four round graphene flakes 
with almost the same diameter and $\sim$5000 atoms. Based on the 
triangular and hexagonal graphene flakes, one might expect that 
the shell structure is independent of the size. Figure \ref{roundfig} 
demonstrates that this is clearly not the case. On the contrary, the 
level structure is very sensitive to the flake diameter. This can be 
understood by inspecting the structure of the low-energy wave functions 
(lower panel in figure \ref{roundfig}). The states above the Fermi 
energy are localized close to the flake edges, and therefore, they 
experience the detailed edge geometry.

The edge of a circular flake comprises not only the simple armchair 
and zigzag segments, but also more complicated parts. Figure 
\ref{roundfig} shows the electron density of the lowest state above 
the Fermi level. In all cases, the electron density is concentrated in 
the zigzag regions. The length and distribution of the zigzag segments 
varies with the flake diameter (size). This causes that the energy of 
the corresponding state is different for each round flake. The same 
argument applies for all the low-energy states since they have marked 
amplitudes at the edges, and it explains the strong size-dependence of 
DOS. The edge states have large degeneracy (or near degeneracy) as seen 
as a large peak at zero energy in the plots of the density of states
in Figs. \ref{dos2}, \ref{hexdos}, and \ref{roundfig}b. This can cause 
spin-polarization in a partially filled case due to the Hund's first rule,
but this is out of the reach of our simple model.

\begin{figure}[!h]
\hfill\includegraphics[width=0.8\textwidth]{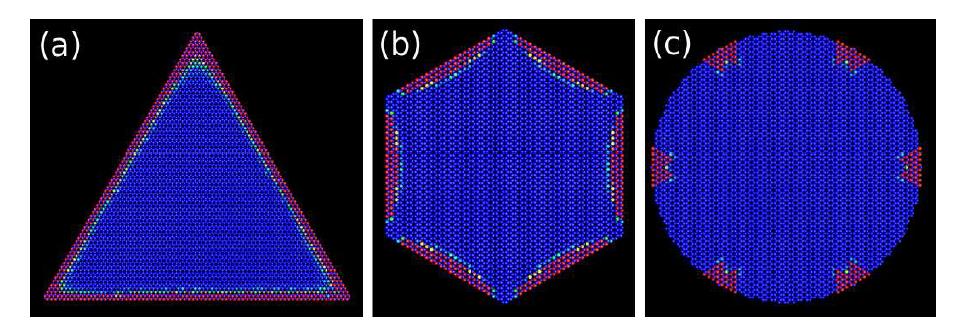}
\caption{Electron density of the edge states (DOS peak at $\epsilon_F$)
for (a) a zigzag triangle (5622 atoms, 72 states), (b) zigzag hexagon 
(3750 atoms, 18 states), and (c) round flake (5118 atoms, 12 states).
Color scale from blue to red, blue corresponds to vanishing density.}
\label{edgestates}
\end{figure}

The edge states are visible in the zigzag triangles and hexagons, and 
they appear as a prominent peak (DOS) at the Fermi energy. For triangles, 
all the edge states have zero energy, i.e. they are exactly at the Fermi 
level. The hexagon edge states are also concentrated at the Fermi energy, 
but they have a small dispersion. The situation is significantly different 
in round flakes due to the fact that the lengths of the zigzag regions are 
very small. This leads to a situation where the electrons become localized 
and their energy increases. Figure \ref{edgestates} shows the total 
electron density of edge states in these three cases: In the case of 
triangles, the edge states bend smoothly around the corners of the triangle, 
but for hexagons they are already pushed out from the corners. The round 
flakes exhibit edge states that are localized in narrow regions and 
penetrate much deeper inside the flake.

\section{Conclusions}

We have computed the electronic structure of triangular, hexagonal 
and round graphene flakes by using a TB method that considers the 
carbon $p_z$ electrons. We observe that the DOS close to the Fermi 
energy $\epsilon_F$ is independent of the size of the triangles and 
armchair-edged hexagons, but depends strongly on the size of the zigzag 
hexagons and round flakes. The triangles with zigzag edge exhibit the
well-known edge states, while the armchair triangles show an additional 
set of ``ghost states" which result in from the interplay between the 
graphene band structure and the plane wave solutions of the wave 
equation. The same ghost states will emerge in any flake of graphene 
that can be constructed from equilateral armchair triangles of the same 
size with additional rows of atoms in the boundaries.

Also hexagons can be constructed with armchair or zigzag edges. In the 
case of the armchair edge, the shell structure is clear and scalable with 
the flake size (cf. triangles). However, for the zigzag edge the shell 
structure of the hexagonal confinement is disturbed by the edge states, 
and the level structure above the Fermi energy depends on the size of 
the hexagon.

For round graphene dots, one might expect the shell structure of a 
circular cavity. However, the low-energy level structure (above 
$\epsilon_F$) is dominated by the edge states that appear in the 
zigzag regions of the edge, and the lengths and distribution of such
segments vary with the flake diameter. Consequently, the level structure 
is very sensitive to the size of the circular graphene flake.

The effect of the edge roughness on shell structure was studied by 
removing a fraction of atoms randomly. In the armchair triangles, the 
shell structure is simple and scalable, and the roughness has only a 
small effect on the low energy states. For the zigzag hexagons, the 
low-energy levels are edge-related, and already a small roughness 
removes the shell structure.

We have obtained our results for free graphene flakes and not considered 
the interaction with substrate or electric leads which evidently could have 
effects of the shell structure. However, transport spectroscopy through
semiconductor quantum dots\cite{tarucha1996} 
have shown that the shell structure calculated for free dots\cite{koskinen1997} 
can indeed be captured with weak connections to leads.
A more direct measurement of the electronic states would be scanning
tunneling microscopy which has already been used to study suspended
graphene\cite{li2008}. It is possible that on a proper surface 
STM spectroscopy could reveal the detailed structures of the electron 
wave functions.

\end{document}